\documentclass[sigconf,10pt]{acmart}
\newif\ifarXiv
\arXivtrue
\ifarXiv
\renewcommand\footnotetextcopyrightpermission[1]{} % removes footnote with conference information in first column
\fi
\usepackage{enumitem}
\usepackage{booktabs}
\usepackage{subfigure}
\usepackage{makecell}

\newif\ifcomm
\commtrue %Comment to turn off comments
\ifcomm
\else
\commfalse
\fi
\ifcomm
\newcommand\jl[1]{\textcolor{red}{JL: #1}}
\newcommand\minlan[1]{\textcolor{blue}{MY: #1}}
\newcommand\MM[1]{\textcolor{green}{MM: #1}}
\newcommand\ga[1]{\textcolor{brown}{GA: #1}}
\newcommand\ran[1]{\textcolor{orange}{Ran: #1}}
\newcommand\sivaram[1]{\textcolor{violet}{SR: #1}}
\newcommand\go[1]{\textcolor{yellow}{GO: #1}}
\else
\newcommand\jl[1]{\textcolor{red}{}}
\newcommand\minlan[1]{\textcolor{blue}{}}
\newcommand\MM[1]{\textcolor{green}{}}
\newcommand\ga[1]{\textcolor{brown}{}}
\newcommand\ran[1]{\textcolor{orange}{}}
\newcommand\sivaram[1]{\textcolor{violet}{}}
\newcommand\go[1]{\textcolor{yellow}{}}
\fi

\setcopyright{rightsretained}
\acmDOI{10.475/123_4}

\acmISBN{978-1-4503-7020-2}

\copyrightyear{2021}
\acmYear{2021}
\setcopyright{acmlicensed}\acmConference[HotNets '21]{The Twentieth ACM Workshop on Hot Topics in Networks}{November 10--12, 2021}{Virtual Event, United Kingdom}
\acmBooktitle{The Twentieth ACM Workshop on Hot Topics in Networks (HotNets '21), November 10--12, 2021, Virtual Event, United Kingdom}
\acmPrice{15.00}
\acmDOI{10.1145/3484266.3487366}
\acmISBN{978-1-4503-9087-3/21/11}

\begin{document}
%%%%%%%%%%%% THIS IS WHERE WE PUT IN THE TITLE AND AUTHORS %%%%%%%%%%%%

\title{Zero-CPU Collection with Direct Telemetry Access}

\author{Jonatan Langlet}
\orcid{0000-0003-0644-6612}
\affiliation{%
  \institution{Queen Mary University of London}
  %\institution{QMUL}
}
%\email{j.langlet@qmul.ac.uk}

\author{Ran Ben-Basat}
\orcid{0000-0003-0196-9190}
\affiliation{%
  \institution{University College London}
  %\institution{UCL}
}
%\email{r.benbasat@ucl.ac.uk}

\author{\hspace*{1mm}{Sivaramakrishnan Ramanathan}}
\affiliation{%
  \institution{\hspace*{-7mm}\mbox{University of Southern California}}
  %\institution{USC}
}
%\email{satyaman@usc.edu}

\author{Gabriele Oliaro}
\affiliation{%
  \institution{Harvard University}
}
%\email{gabriele\_oliaro@college.harvard.edu}

\author{\mbox{\hspace*{-4mm}Michael Mitzenmacher}}
%\author{Michael Mitzenmacher}
\orcid{0000-0001-5430-5457}
\affiliation{%
  \institution{Harvard University}
}
%\email{michaelm@eecs.harvard.edu}

\author{Minlan Yu}
\orcid{0000-0002-2381-0212}
\affiliation{%
  \institution{Harvard University}
}
%\email{minlanyu@g.harvard.edu}

\author{\hspace*{8mm}Gianni Antichi}
\orcid{0000-0002-6063-4975}
\affiliation{%
  \institution{\hspace*{-2mm}\mbox{Queen Mary University of London}}
  %\institution{QMUL}
}
%\email{g.antichi@qmul.ac.uk}

%%%%%%%%%%%%%  ABSTRACT MUST PRECEDE MAKETITLE in this style %%%%%%%%%%%%%%
\begin{abstract}
Programmable switches are driving a massive increase in fine-grained measurements. This puts significant pressure on telemetry collectors that have to process reports from many switches. Past research acknowledged this problem by either improving collectors' stack performance or by limiting the amount of data sent from switches.
In this paper, we take a different and radical approach: switches are responsible for directly inserting queryable telemetry data into the collectors' memory, bypassing their CPU, and thereby improving their collection scalability.
We propose to use a method we call \emph{direct telemetry access}, where switches jointly write telemetry reports directly into the same collector's memory region, without coordination. Our solution, DART, is probabilistic, trading memory redundancy and query success probability for CPU resources at collectors. 
We prototype DART using commodity hardware such as P4 switches and RDMA NICs and show that we get high query success rates with a reasonable memory overhead.
For example, we can collect INT path tracing information on a fat tree topology without a collector's CPU involvement while achieving 99.9\% query success probability and using just 300 bytes per flow.
\end{abstract}

\maketitle

\renewcommand{\shortauthors}{Langlet et al.}

\vspace*{-1mm}
\section{Introduction}\label{sec:intro}

Network telemetry is an integral function in modern data centers~\cite{ben2020pint,huang2020omnimon,intelDeepInsight,khandelwal2019confluo,kim2015band,van2017towards,yu2019network,zhou2020hypersight,zhou2020flow}. This is fostered by the rise of programmable switches~\cite{trident,intelTofino,mellanox_spectrum} that allows monitoring of network traffic in real time at high granularities. 
Such granular telemetry is essential both for advanced network operations~\cite{li2019hpcc,alizadeh2014conga,heller2010elastictree} and troubleshooting~\cite{guo2015pingmesh,intelDeepInsight,tammana2018distributed}.

Telemetry systems are built around centralized collection of network-wide reports~\cite{aristaTelemetry,ciscoMDT,huaweiTelemetry,intelDeepInsight,juniperTelemetry}.
However, the growing telemetry volume poses a new challenge: it is increasingly hard to build collectors that can process reports (i.e., telemetry data) from many switches~\cite{khandelwal2019confluo,van2018intcollector}.
For example, production datacenter networks can comprise hundreds of thousands of switches~\cite{guo2015pingmesh}, each generating up to millions of reports per second~\cite{zhou2020flow}, requiring thousands of CPU cores just for real-time data collection (\S\ref{sec:motivation}).
Existing research boosts collectors' scalability by improving their network stacks~\cite{khandelwal2019confluo,van2018intcollector} or by preprocessing~\cite{li2020concerto} and filtering data at the switches~\cite{intelINT,kuvcera2020enabling,vestin2019programmable,zhou2020flow}. 
Our insight, however, is that the main bottleneck of collectors is their inability to quickly insert incoming \mbox{reports in queryable data structures (\S\ref{sec:motivation}).} 

To overcome this issue, we propose a method we call \emph{direct telemetry access} where switches write their reports directly into a collector's memory. Our solution, DART (Distributed Aggregation of Rich Telemetry), allows switches to jointly insert queryable telemetry data without any involvement of the collector's CPU or inter-switch communication.
DART uses RDMA (Remote Direct Memory Access)~\cite{rdma}, a technology available on many network cards~\cite{intelE810,mellanoxConnectx6,xilinxERNIC} that can perform hundreds of millions of memory writes per second~\cite{mellanoxConnectx6}, which is significantly faster than what even the most high-performing CPU-based telemetry collectors achieve~\cite{khandelwal2019confluo}. 
Generating RDMA instructions directly from switches is possible~\cite{kim2020tea}, but it also raises several challenges when used for telemetry collection: 
(1) how to directly write in the collector memory in such a way that the data is then easily queryable? 
(2) how to optimize data organization inside the collector in the presence of hundreds of thousands concurrent telemetry reporters? 
(3) how to make the system robust to telemetry report losses while keeping limited statefulness at switches?
To address these challenges, we designed a solution where switches decide the location in collectors' memory to write the reports. This is achieved using global hash functions that create a stateless mapping between the information to be reported (i.e., telemetry keys and data), and the memory addresses at collectors. 
Such a mapping allows collectors to determine where the information relevant to a query is stored as the same address mapping can be used to retrieve data. 
However, different switches might write to the same memory location, thus potentially deleting useful telemetry data. To overcome this, DART switches write the same report to multiple memory addresses. %\jl{this reads strange? just me? "with DART"}
This trades the amount of memory needed and the query success rate for CPU resources at collectors (which are limited by the slowdown of Moore's law~\cite{esmaeilzadeh11,theRiseOfDarkSilicon}).
These combined techniques manage to deliver both \mbox{coordination-free collection, and collision robustness.}

    We discuss DART's design~(\S\ref{sec:design}), provide a theoretical analysis~(\S\ref{sec:theory}), and confirm the efficiency with simulations~(\S\ref{sec:evaluation}) where we use INT path tracing carried on a 5-hop fat-tree topology as an example. Here, DART requires as little as 300 bytes per flow to achieve a 99.9\% success probability.
    We show that DART is efficiently implementable in commodity P4 switches~(\S\ref{sec:implementation}), and \mbox{discuss future directions~(\S\ref{sec:discussion}).}
    
    \textbf{Our main contributions are:}
    
    \begin{itemize}[leftmargin=0.12in,nosep]
        \item We make the case for adopting a solution that does not use CPU at collectors to handle incoming telemetry data~(\S\ref{sec:motivation}).
        \item We propose a method called \textit{direct telemetry access} that allows switches to jointly insert queryable telemetry data into collectors' memory~(\S\ref{sec:design}).
        \item We set the basis for \textit{direct telemetry access} theory demonstrating how it is possible to have provable query success rate given collector's memory availability~(\S\ref{sec:theory}).
        \item We demonstrate the feasibility of our approach using state-of-the-art programmable switches~(\S\ref{sec:implementation}).
    \end{itemize}
\vspace*{-2mm}
\section{Motivation}\label{sec:motivation}
    Collectors play an important role in network telemetry systems: they receive telemetry reports and store the information in internal data structures to answer network-wide queries. 
    One key challenge is to ensure that this process is scalable as a datacenter network can comprise hundreds of thousands of switches~\cite{guo2015pingmesh}, some potentially handling up to millions of traffic flows~\cite{roy2015inside}.
    For example, a non-sampled INT telemetry system requires the collection of telemetry data from \emph{every single packet}, which would result in an excessive amount of reports. Because of this, event detection is typically implemented at switches in an effort to send reports to a collector only when things change~\cite{intelINT}. 
    This helps in reducing the rate of switch-to-collector communication down to a few million telemetry reports per second per switch~\cite{zhou2020flow}. 
    Still, telemetry collection costs are high, and the main reason we identified is that \mbox{the collectors' CPU is the main bottleneck.}

\begin{figure}
    \vspace*{-2mm}
        \centering     %%% not \center
        \subfigure[Pure DPDK Packet I/O cost]{\label{fig:motivation_IO}\includegraphics[width=0.540\linewidth]{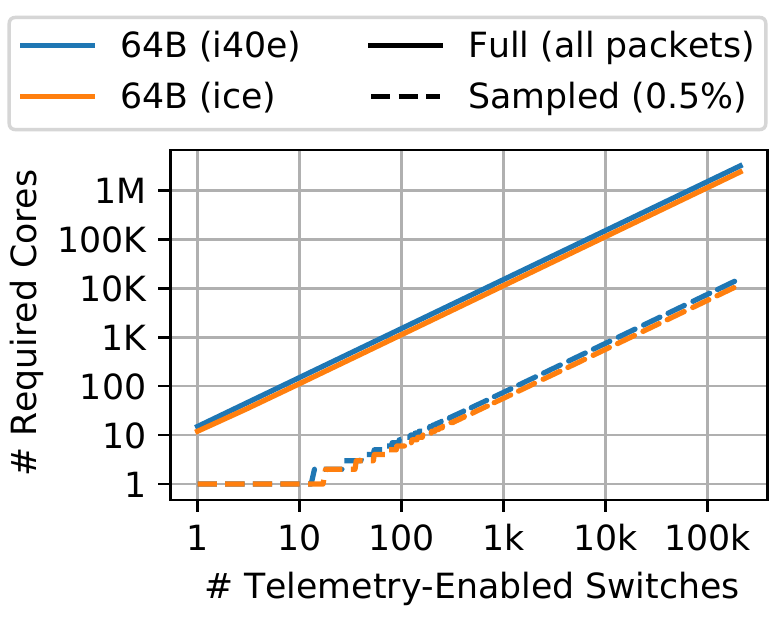}}
        \subfigure[Storage overhead]{\label{fig:motivation_storage}\includegraphics[width=0.435\linewidth]{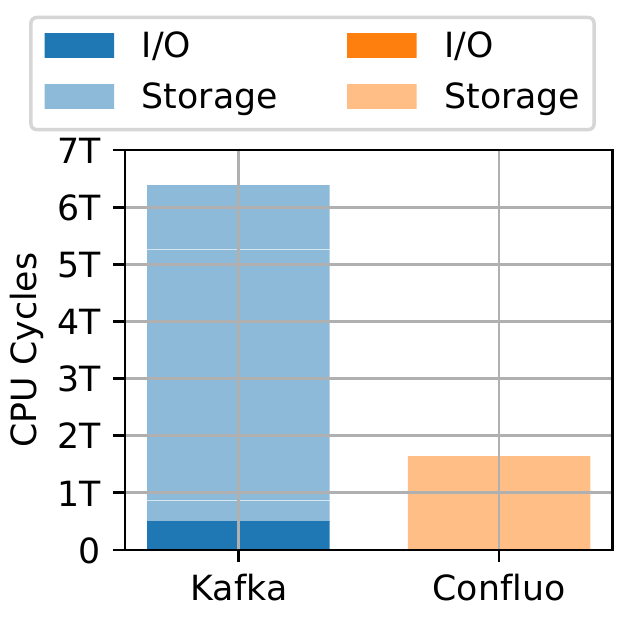}}
        \vspace{-0.22in}
        \caption{Telemetry report packet I/O is already expensive, and collector functionality requires significant additional work. I/O performance and sampling in (a) are based on official DPDK PMD performance numbers~\protect\cite{DPDKPerf} and generated events per second in 6.5Tbps switches~\protect\cite{zhou2020flow}. (b) shows the CPU cycles breakdown to storage and I/O in existing collectors.}
        \vspace*{-0.198202in} 
    \end{figure}

\textbf{CPU-based packet I/O is too slow.}
    Figure~\ref{fig:motivation_IO} shows the number of CPU cores required by a collector when using the DPDK PMD (Poll Mode Driver), a state-of-the-art kernel bypass approach, to \emph{just receive} telemetry report packets at 64 and 128 bytes \emph{including the headers}\footnote{We assume that reports are not batched together into fewer packets for reduced I/O overhead. However, this would not remove the cost of processing the \mbox{reports and inserting them into storage, which is the most costly step.}}
    Even normal-sized data centers, comprising 10K switches, would require a collection cluster containing thousands of CPU cores dedicated to simple packet I/O.
    However, further processing is then essential to \mbox{ensure telemetry data insertion into queryable storage.}

\textbf{CPU-based telemetry storage is slower.}
    Figure~\ref{fig:motivation_storage} shows the number of CPU cycles required for packet I/O and insertion of telemetry reports into storage.
    We used two state-of-the-art solutions to store the contents of telemetry reports: Apache Kafka~\cite{garg2013apache} with socket-based packet I/O, and  Confluo~\cite{khandelwal2019confluo} with DPDK-based packet I/O.  
    We uniformly generate two different report types that are 64 and 128 bytes\footnote{A 64 or 128 bytes report would consist of 36 bytes and 100 bytes of report data (without 28 bytes of header). For instance, a 64 bytes packet could answer one INT query, storing 32-bits per hop across 9 hops in the network.}.
    Socket-based packet I/O is inefficient, requiring 504 billion CPU cycles for processing 100 million reports, with 11.5x as many additional cycles required by Kafka.
    CPU overhead from packet I/O is significantly reduced by the DPDK PMD, which requires only 14 billion CPU cycles for the same number of reports (i.e. $2.7\%$ as much work as sockets). 
    However, as visualized in Figure~\ref{fig:motivation_IO}, this is still very expensive at large scales.
    The actual insertion of the telemetry data into queryable storage through Confluo requires an astounding \emph{114x as many CPU cycles} as the costly packet I/O.
    
\textbf{Direct telemetry access to the rescue.} 
    To eliminate the processing bottleneck at collectors, we designed a solution where switches are responsible for directly inserting queryable telemetry data into logically centralized memory.
    We show how this can be achieved using commodity hardware such as P4 programmable switches and RDMA NICs. Current RDMA-capable network cards are capable of processing more than 200 million messages per second~\cite{mellanoxConnectx6}, which is significantly faster than CPU-based telemetry collectors~\cite{khandelwal2019confluo}. 
    Our solution is not restricted to the RDMA protocol, and we discuss in Section~\ref{sec:discussion} how smartNICs can be leveraged to build a new protocol tailored to direct telemetry access \mbox{for significant optimizations.}

\section{Design}\label{sec:design}
    \begin{figure}
        \centering
        \includegraphics[width=1.0\linewidth]{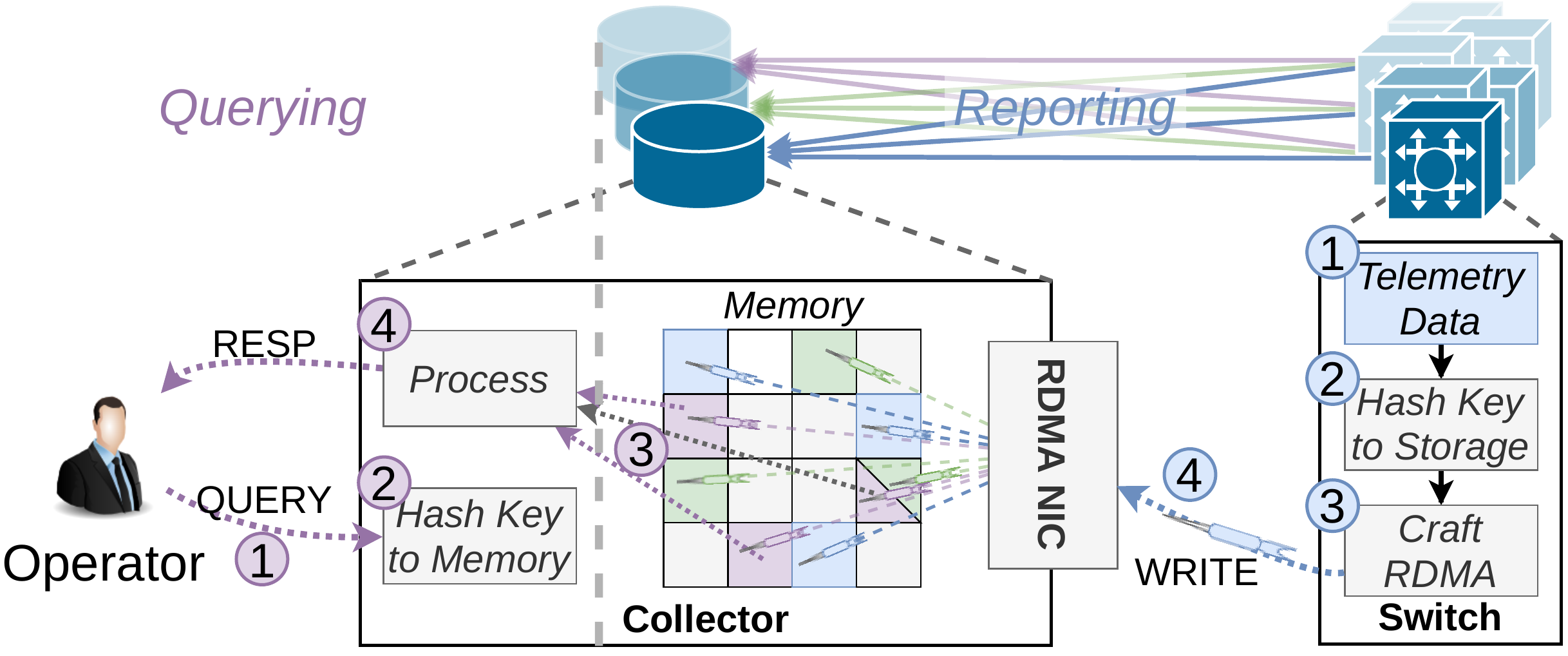}
        \vspace{-0.3in}
        \caption{An architectural overview of DART.}
        \label{fig:overview}
        \vspace{-0.30in}
    \end{figure}

    In DART, switches insert telemetry data directly into the collectors' memory at a specific address using RDMA calls. 
    In Figure~\ref{fig:overview}, we show the architectural overview of DART, constituting two main components: telemetry \emph{reporting}~(\S\ref{sec:design_reporting}), shown on the right side of the figure, and \emph{querying}~(\S\ref{sec:design_querying}), shown on the left side. 
    The former is the process of pushing the network state and measurements from switches directly into collectors' memory, organized as a key-value store. 
    The latter refers to the operator's ability to run key-based queries against the stored telemetry data.
    Both of these functions are delivered without requiring any coordination or communication between individual switches or end-hosts, further reducing the \mbox{overall telemetry overhead and complexity.}
    
    DART assumes the telemetry data is readily available at switches, and thus it does not place any specific restriction on the underlying measurement framework, as shown in Table~\ref{tab:support_examples}.
    Because of this, the key-value store semantics may depend on the specific telemetry techniques used at switches.
    For example, for INT~\cite{INTSpec}, each switch writes its telemetry data into packets and only the last hop pushes the information to the collector. Here, the key will be the \textit{<Flow 5-tuple>}.
    In contrast, when DART is used with INT working in postcard mode, where each switch reports data, the key will be the \mbox{concatenation of \textit{<Flow 5-tuple>} and the \textit{<switchID>}.}
    
    \vspace*{-1mm}
    \subsection{Reporting}\label{sec:design_reporting}
        One might be tempted to design collision-free key-to-address mappings using on-switch memory, combined with dedicated per-switch regions in the centralized telemetry storage. We believe this approach is unfeasible due to the switches' limited memory, unable to store the high number of new telemetry keys constantly appearing across the entire network, as old keys become irrelevant. Indeed, there is no support for dynamic memory allocations, which has led the community to adopt probabilistic data structures to approximate per-key statefulness in switches~\cite{basat2020designing,basat2020routing,liu2016one}. Second, keeping dedicated per-switch regions in the collector's memory leads to inefficiencies due to uneven switch telemetry data generation rates. Further, operators need to know which switch wrote \mbox{the telemetry data for it to be efficiently queryable.}
        
        %2
        DART uses global hash functions to create a stateless mapping between telemetry keys and memory addresses at one or more logically centralized collectors where the data should be written.
        The same mapping can be calculated by the operator for retrieving the results of a query.
        However, using stateless mapping raises an important issue -- different keys can hash into the same memory address, resulting in overwritten telemetry data due to these collisions.
        To address this challenge, DART uses $N$ independent hash functions to map each key into $N$ different storage locations, where duplicate \mbox{entries of the corresponding telemetry data are stored.}
        
        \begin{table}[]
            \centering\renewcommand\cellalign{lc}
            \setcellgapes{0pt}\makegapedcells
            \resizebox{\linewidth}{!}{% <------ Don't forget this %
            \begin{tabular}{@{}llll@{}}
                \toprule
                \begin{tabular}[c]{@{}l@{}}\textbf{Backend}\end{tabular} & \textbf{Key(s)} & \textbf{Data} \\ 
                \midrule
                \noalign{\global\arrayrulewidth=0.001mm}
                \begin{tabular}[c]{@{}l@{}}In-band~\cite{ben2020pint,INTSpec,jeyakumar2014millions,kim2015band}\end{tabular} & \begin{tabular}[c]{@{}l@{}}Flow 5-tuple\end{tabular} & \begin{tabular}[c]{@{}l@{}}Packet-carried data\end{tabular} \\\hline
                \begin{tabular}[c]{@{}l@{}}Postcards~\cite{INTSpec,handigol14}\end{tabular} & \begin{tabular}[c]{@{}l@{}}SwitchID, \\ Flow 5-tuple\end{tabular} & \begin{tabular}[c]{@{}l@{}}Local\\ measurement\end{tabular} \\\hline
                \begin{tabular}[c]{@{}l@{}}Query-based\\ mirroring~\cite{zhu2015packet}\end{tabular} & QueryID & Query answer \\\hline
                \begin{tabular}[c]{@{}l@{}}Trace analysis~\cite{rasley2014planck,yu2019dshark}\end{tabular} & \textit{various} & Analysis output \\\hline %Very vague, but trace analysis can do basically anything
                Flow anomalies~\cite{zhou2020flow} & \begin{tabular}[c]{@{}l@{}}Flow 5-tuple,\\ Anomaly ID\end{tabular} & Time, \textit{event-specific} \\\hline
                Network failures~\cite{guo2015pingmesh} & Failure ID, \textit{location} & Time, \textit{debug info} \\%\hline
                \bottomrule
            \end{tabular}
            }
            %\vspace{-0.15in}
            \caption{Examples of measuring techniques mapped into the DART key-value collection structure.}
            \label{tab:support_examples}
            \vspace{-0.4in}
        \end{table}  
    
        Telemetry reports are sent from switches as one-sided RDMA-WRITE packets towards one of the collectors, with the chosen memory address in the RDMA header. Through hashing, switches determine the collector and memory location for $N$ copies of the telemetry report. Switches craft RDMA-based reports based on loaded lookup-table entries that map the selected collector to essential RDMA information.
        An RDMA-capable network card at the collector parses the RDMA report and writes the payload directly \mbox{to memory, making it available for operator queries.}
        
        RDMA does not support writing a payload into several memory addresses at once, instead requiring several packets containing one memory instruction each. Currently, DART-enabled switches rely multiple redundant telemetry reports generated to fill all the $N$ slots allocated to a key. 
        However, further research into SmartNIC-enabled RDMA extensions could remove this requirement~(\S\ref{sec:discussion}).
        
        At the collector, to reduce the memory occupied, instead of storing the key, DART makes use of a small checksum of the key to simplify detection of overwritten data.
        While querying, values from the $N$ locations that do not match the attached checksum can be discarded. The impact of checksum collisions is discussed in sections \ref{sec:theory} and \ref{sec:eval_correctness}.
        
        Distributing the $N$ copies of per-key telemetry data across $N$ physical collectors could improve the system resiliency, at the cost of potentially reduced querying speed.
        In DART's current design we ensure that data duplicates for any one key are held at a single collector, thereby enabling operator queries to be executed locally on the collector CPU without \mbox{requiring inter-collector communications for data transfer.}
        
    \subsection{Querying}\label{sec:design_querying}
        Queries are performed in four main steps, as seen in the left side of Figure~\ref{fig:overview}.
        First, an operator requests the results for a telemetry query, which it forwards to the relevant collector. 
        DART hashes the query key to retrieve the collector ID, and then uses a lookup table to convert the collector ID into the collector which holds the telemetry data.
        DART then hashes the key into $N$ memory addresses at the collector which holds the relevant telemetry data and extracts this data.
        Finally, DART uses the key-checksum to discard the invalid telemetry \mbox{data and returns the result of the query to the operator.}

\vspace*{-1mm}
\section{Theoretical Analysis}\label{sec:theory}
Because we treat the RDMA memory as a large key-value hash table where only checksums of keys are stored and values may be overwritten over time, we must consider the possibility that when we make a query, we are unable to return an answer, or we may return an incorrect answer.  
We call the case where we have no answer to return an {\em empty return}, and the case where we return an incorrect answer a {\em return error}.
The probability of an empty return or a return error depends on the parameters of the system, and on the method we choose to determine the return value.  Below we present some of the possible tradeoffs and some mathematical analysis; we leave \mbox{further results and discussions for the full paper.}

Let us first consider a simple example.  When a write occurs for a key-value pair, in the hash table $N$ copies of the $b$-bit key checksum and the value are stored at random locations. 
We assume the checksum is uniformly distributed for any given key throughout our analysis.
When a read occurs, let us suppose we return a value if there is only a single value amongst the $N$ memory locations matching that checksum.  (The value could occur multiple times, of course.)

An empty return can occur, for example, if when we search the $N$ locations for a key, none of them have the right checksum.  
That is, all $N$ copies of the key have been overwritten, and none of the $N$ locations currently hold another key with the same checksum.
To analyze this case, let us consider the following scenario.  
Suppose that we have $M$ memory cells total, and that there are $K = \alpha M$ updates of {\em distinct} keys between when our query key $q$ was last written, and when we are making a query for its values.  
We can use the Poisson approximation for the binomial (as is standard in these types of analyses and accurate for even reasonably large $M$, $N$, $K$; see, for example, \cite{broder2004network,mitzenmacher2017probability}).
Using such approximations, the probability that any one of the $N$ locations is overwritten is given by $(1-e^{-KN/M})$, and that all of them are overwritten is $(1-e^{-KN/M})^N$.  
The probability that all of them are
overwritten and \mbox{the key checksum is not found is approximated by} $$(1-e^{-KN/M})^N \cdot (1-2^{-b})^N = (1-e^{-\alpha N})^N \cdot (1-2^{-b})^N.$$

We would also get an empty return if the $N$ cells contained two or more distinct  values with the same correct checksum.

\vspace{1mm}\noindent
This probability is lower bounded by
{\small{
$$ \sum_{j=1}^{N-1} {N \choose j} (1-e^{-\alpha N })^j e^{-\alpha N(N-j)} (1 - (1-2^{-b})^j)\quad ,$$ 
}}
and upper bounded by  
{\small{
\begin{multline*}
    \Big( \sum_{j=1}^{N-1} {N \choose j} (1-e^{-\alpha N})^j e^{-\alpha N(N-j)} (1 - (1-2^{-b})^j) \Big)\\
    \qquad\qquad + (1-e^{- \alpha N})^N (1 - (1-2^{-b})^N - N\cdot 2^{-b}(1-2^{-b})^{N-1}).
\end{multline*}
}} 
\noindent The first summation is the probability at least one of the original $N$ locations is not overwritten, but at least one overwritten location gets the same checksum.  (We pessimistically assume it obtains a different value.)
The second expression adds a term for when all original values are overwritten and two or more obtain the same checksum. Note that we need to give bounds as values in overwritten locations \mbox{may or may not be the same.}

We could have a return error if all $N$ copies of the original key are overwritten and one or more of those cells are overwritten with the same checksum and same (incorrect) value.  
This probability is lower bounded by $$(1-e^{-\alpha N})^N N 2^{-b} (1-2^{-b})^{N-1},$$  which is the probability that all of the original locations are overwritten and a single overwriting key obtains the checksum, and upper bounded by   $$(1-e^{-\alpha N})^N (1 - (1- 2^{-b})^N),$$  the probability that the original locations are overwritten \mbox{and at least one overwriting key obtains the checksum.}

There are many ways to modify the configuration or return method to lower the empty returns and/or return errors, at the cost of more computation and/or more memory.  
The most natural is to simply use a larger checksum; we suggest 32 bits should be appropriate for many situations. 
However, we note that at ``Internet scale'' rare events will occur, even matching of 32-bit checksums, and so this should be considered when utilizing DART information.
One can also use a ``plurality vote'' if more than one value appears for the queried checksum;  additionally one can require that a checksum/value pair occur at least twice among the $N$ values before being returned.  (Note that, for example, requiring consensus of two values can be decided on a per query basis without changing anything else; one can decide for specific queries whether to trade off empty returns and return errors this way.) 
Additional ideas from coding theory \cite{goodrich2011invertible,lipton1994new}, including using different checksums for each location or XORing each value with a pseudorandom value, could also be applied.  
As a default, we suggest a 32-bit checksum and a ``plurality vote.'' We \mbox{describe related results in our evaluation~(\S\ref{sec:evaluation}).}

\section{Preliminary Evaluation}\label{sec:evaluation}
    RDMA is well known to deliver high throughput memory operations, and this section focus on evaluating the DART algorithm and data structure.% for telemetry collection.
    We show through in-depth simulations that DART is effective with little redundancy of $N=2$ (\S\ref{sec:eval_redundancy}) and has a high query success rate of 99.9\% at the collector~(\S\ref{sec:eval_data_queryability}) with high accuracy~(\S\ref{sec:eval_correctness}).
    
    \subsection{Effectiveness of DART Redundancy}\label{sec:eval_redundancy}
        \begin{figure}[ht]
            \vspace*{-5mm}
            \centering
            \includegraphics[width=\linewidth]{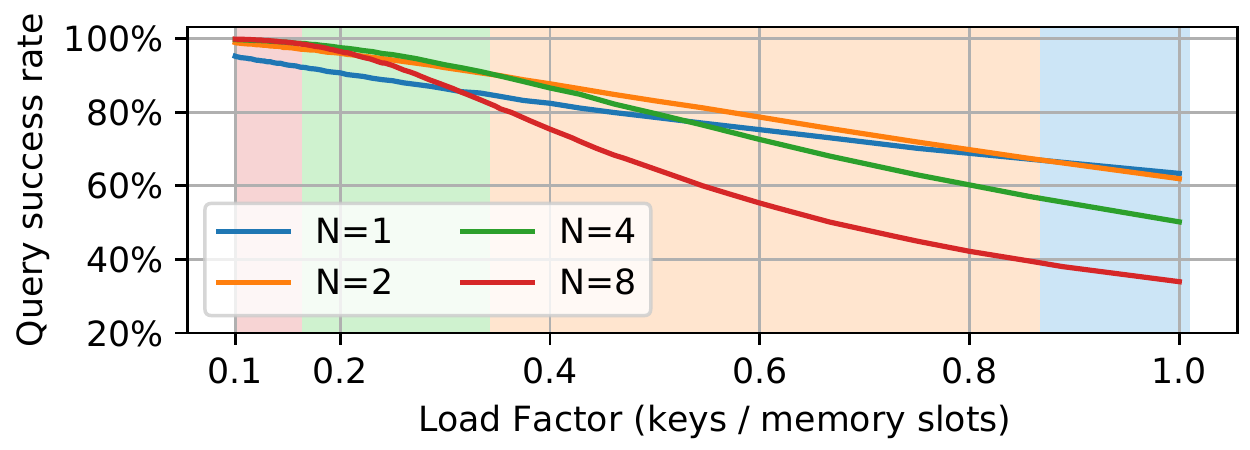}
            %\includesvg[width=\linewidth]{Figures/query_success_rate_N.svg}
            \vspace{-0.3in}
            \caption{Average query success rates in DART, depending on the collector load and the number of addresses per key ($N$). The background color indicates optimal $N$ in each interval.}
            \label{fig:success_rate_N}
            \vspace{-0.1in}
        \end{figure}
        The probabilistic nature of DART cannot guarantee queryability on a given reported key.
        We show in Figure~\ref{fig:success_rate_N} how the query success rate depends on the load factor (i.e., the total number of telemetry keys over available memory addresses), and the number of memory addresses that each key can write to.
        There is a clear efficiency improvement by having keys write to $N>1$ memory addresses when the storage load factor is in reasonable intervals.
        We also note how simulations adhere to the aforementioned theory in Section~\ref{sec:theory} regarding the impact of multiple addresses per key, and the background color in Figure~\ref{fig:success_rate_N} indicate which number of addresses per key ($N$) delivered the highest key queryability in each interval.
        
        The RDMA standard requires multiple packets with a single write instruction each, with SmartNICs showing promise to circumvent this limitation (\S\ref{sec:discussion}) by batching them together.
        Thus, a practical RDMA-based DART implementation might benefit from a reduction in $N$, balancing network overheads against the marginal queryability improvements gained from the increased data redundancy. $N=2$ appears to be a generally good compromise, showing great queryability improvements over $N=1$.
        We conclude that dynamically adjusting $N$ as the load fluctuates could improve queryability and efficiency, \mbox{and leave finding a good mechanism as future work.}

    \subsection{Data Queryability}\label{sec:eval_data_queryability}
        \begin{figure}[ht]
        \vspace*{-4mm}
            \centering
            \includegraphics[width=\linewidth]{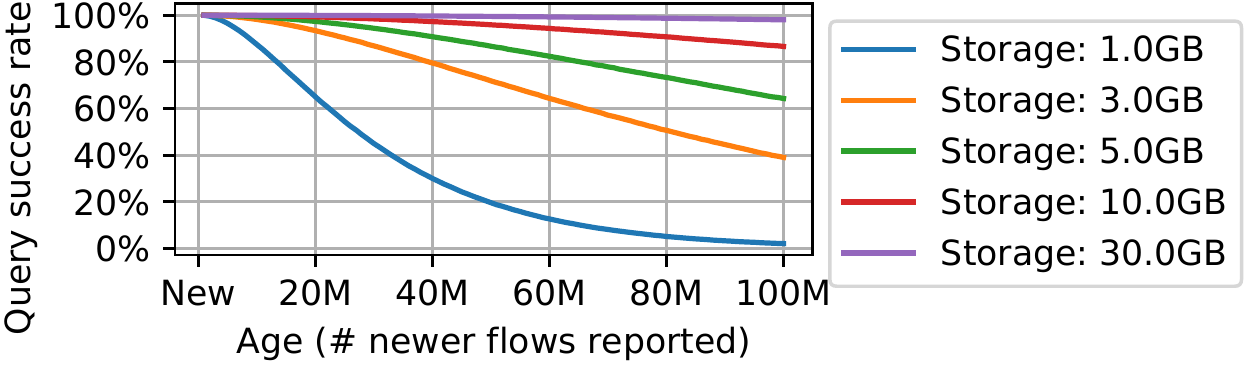}  
            \vspace{-0.3in}
            \caption{Telemetry data aging in DART, showing INT 5-hop path tracing queryability of 100 million flows at various storage sizes, using 160-bit values with 32-bit checksums, with redundancy $N=2$.}
            \label{fig:data_aging}
            \vspace{-0.1in}
        \end{figure}
        The hash-based address selection in DART results in address collisions between keys, as they compete over limited allocated memory at various sizes of the DART data storage, resulting in old data being aged out of memory.

        Figure~\ref{fig:data_aging} shows the queryability of reported INT path tracing data at various storage sizes and report ages.
        As expected, allocating enough collector memory is essential for ensuring a high data queryability, with smaller storage sizes resulting in a faster aging-out of data.
        For example, when 100 million flows share just 3GB (i.e., 30B storage per flow path), we see how the average queryability is $71.4\%$ across all 100 million flows; with a steep decline to $39.0\%$ for the oldest reports, which almost exactly matches the theoretically predicted $38.7\%$ from Section~\ref{sec:theory}. %Theory remains almost perfect with N=2
        However raising the storage capacity to 30GB significantly increases the average data queryability to $99.3\%$ across all 100 million flows; equivalent experiments with redundancy $N=4$ further improves the data queryability to $99.9\%$.
        We also note how the number of tracked flow paths at a given probability increases \mbox{linearly alongside the amount of allocated storage memory.}
        
        \subsubsection{Practical Considerations}
            The ability to run queries on historical data, for example to troubleshoot a previous outage, is important.
            Writing directly to memory is essential for allowing line-rate report ingestion, but fails to scale to the sizes that would be needed for storing historical network-wide measurements.
            A solution can be to utilize DRAM for temporary epoch-based storage of telemetry data, combined with periodical transfer of data into a larger (and much slower) persistent storage where historical queries can be answered.
            We leave the design details as future work.
            
    \subsection{Query Answer Correctness}\label{sec:eval_correctness}
        \begin{figure}[ht]
            \vspace*{-3mm}
            \centering
            \includegraphics[width=\linewidth]{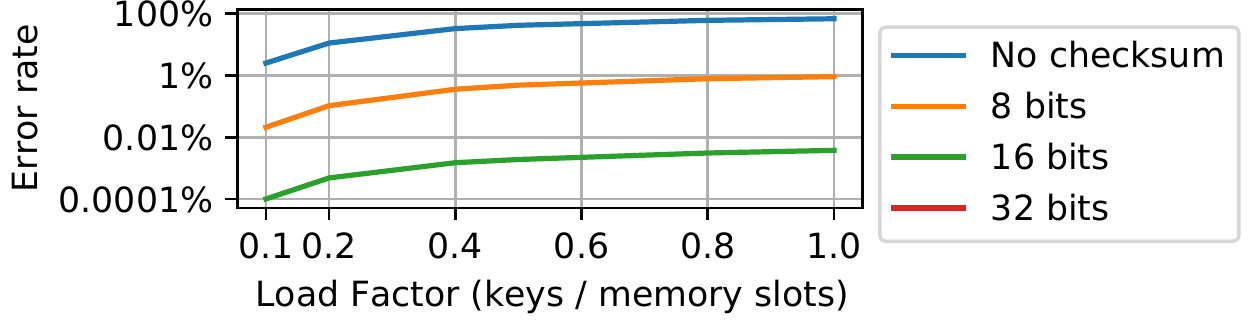}  
            \vspace{-0.37in}
            \caption{The probability of returning the wrong \mbox{answer, due to address and checksum collisions.}}
            \label{fig:eval_correctness}
            \vspace{-0.13in}
        \end{figure}
        There is a theoretical risk of DART returning incorrect query results, as discussed in section~\ref{sec:theory}.
        Figure~\ref{fig:eval_correctness} shows results after extensive tests, where multiple simulations of 100M keys have been performed at various storage sizes in an attempt to recreate the theoretically predicted incorrectness.
        These results clearly show the impact from having key-based checksums included in the DART data structure, with increased lengths greatly reducing the risk of errors.
        Our simulations with 32-bit key-checksums fail to reproduce return-error \mbox{cases, due to their very low probability.}%, even while the data structure is under immense load.
\vspace*{-2mm}
\section{Prototype Implementation}\label{sec:implementation}
\vspace*{-1mm}
    We implemented the switch component of DART in around 1K lines of P4\_16~\cite{bosshart2014p4}, compiled through P4 Studio~\cite{p4studio} for the Tofino ASIC~\cite{intelTofino}, together with 150 lines of Python to handle the switch control plane.
    The implementation is oblivious to the specific monitoring technology. 
    When telemetry data has to be reported, an I2E mirror is triggered, injecting a truncated packet clone into the egress pipeline.
    The packet carries the raw telemetry data together with the corresponding key, \mbox{and is used as the base for crafting a DART report.}
    
    The Tofino-native random number generator calculates $n \in [0,N-1]$ to determine which of the $N$ per-key storage locations to use during report generation. Then, the CRC extern maps ($n$, key) into the corresponding collector ID and memory address.
    The global collector lookup table is a match-action table, and maps the collector ID to specific server information required for crafting RoCEv2~\cite{rocev2} headers.
    The RoCEv2 invariant-CRC (iCRC) checksum is generated by the native CRC extern.
    A register array is used to store per-collector RoCEv2 Packet Sequence Number (PSN) counters. 
    Our prototype requires about 20 bytes of on-switch SRAM per-collector for storing metadata, allowing support for tens of thousands of collectors \mbox{without impacting the pipeline complexity.}
\vspace*{-2mm}
\section{Discussion}\label{sec:discussion}
\vspace*{-1mm}

\textbf{Using standard RDMA calls.} We explored the benefits of using just \emph{RDMA write} calls. However, the RDMA protocol supports additional operations: \emph{Fetch \& Add}, and \emph{Compare \& Swap}. The former increments a value at a specified address by a given amount. The latter compares a value at a specified address with a given value: if they are equal, another specified value will be stored at the address.
Both operations can enable more complex telemetry data structures, possibly improving on query richness and memory efficiency. For example, Fetch \& Add can be used to implement flow-counters directly in collectors' memory (saving resources at switches) or to perform network-wide aggregation of sketches. Compare \& Swap can be used to create more complex storage methods. For example, for $N=2$ hashes and an initially empty table, we can use an RDMA write with one hash and Compare \& Swap with another (writing to a second slot only if it is empty), which \mbox{simulations show can potentially improve queryability.}

\textbf{Building new direct telemetry access protocols.} An RDMA call is ultimately a DMA operation from the server's NIC to its main memory. Programmable NICs can enable new RDMA primitives~\cite{amaro2020remote,sidler2020strom}, even requiring multiple DMA calls per-packet~\cite{amaro2020remote}. Similarly, this can open new opportunities to rethink how direct telemetry access is implemented. For example, it would be possible to design a new primitive for inserting the same data into multiple memory addresses. This would significantly reduce the network overheads of our current system which is restricted by RDMA and thus allows only a single memory write per packet. Moreover, it could be possible to design a new key-value store that is more resilient to collisions, with the NIC preemptively \mbox{managing overwrites in some manner.}
\vspace*{-0mm}
\section{Related Works}
\vspace*{-0mm}

\noindent\textbf{Telemetry.}
Traditional techniques have looked into periodically collecting telemetry data~\cite{guo2015pingmesh,hare2011simple,handigol14}. Even though these techniques generate coarse-grained data, they can be significant given the large scale of today's networks. The rise in programmable switches has enabled fine-grained telemetry techniques that generate a lot more data~\cite{tammana2018distributed,INTSpec,ben2020pint,zhou2020flow,zhu2015packet, gupta2018sonata}.
Irrespective of the techniques, collection is identified to be the main bottleneck in network-wide telemetry, and previous works focus on either optimizing the collector stack performance~\cite{khandelwal2019confluo,van2018intcollector}, or reducing the load through offloaded pre-processing~\cite{li2020concerto} and in-network filtering~\cite{intelINT,kuvcera2020enabling,vestin2019programmable,zhou2020flow}.
To our knowledge, all current collection solutions are CPU-based and thus have the same fundamental performance bottleneck.
An alternative approach is letting end-hosts assist in network-wide telemetry~\cite{tammana2018distributed,huang2020omnimon}, which unfortunately requires \mbox{significant investments and infrastructure changes.}

\smallskip
\vbox{\noindent\textbf{Switch-generated RDMA calls}
Recent work has shown that programmable switches can perform RDMA calls~\cite{kim2020tea}, and that programmable network cards are capable of expanding upon {RDMA with new and customized primitives~\cite{amaro2020remote}.}
Especially FPGA network cards show great promise for high-speed custom RDMA~\cite{sidler2020strom,mansour2019fpga}.\\
\smallskip
\textbf{Acknowledgments.} We thank the anonymous reviewers. This work is sponsored in part by the UK EPSRC project EP/T007206/1 and NSF grants CNS-2107078, CCF-2101140, CCF-1563710, and DMS-2023528.
%Michael Mitzenmacher was supported in part by NSF grants CCF-2101140, CNS-2107078, CCF-1563710, and DMS-2023528.\ran{This looks weird. Either add MM's grants into the global ``work sponsored by'' list or specify who was sponsored by which grant.}
}

\clearpage

\bibliographystyle{ACM-Reference-Format.bst} 
\bibliography{references}


\begin{thebibliography}{57}

%%% the Web address.

\ifx \showCODEN    \undefined \def \showCODEN     #1{\unskip}     \fi
\ifx \showDOI      \undefined \def \showDOI       #1{#1}\fi
\ifx \showISBNx    \undefined \def \showISBNx     #1{\unskip}     \fi
\ifx \showISBNxiii \undefined \def \showISBNxiii  #1{\unskip}     \fi
\ifx \showISSN     \undefined \def \showISSN      #1{\unskip}     \fi
\ifx \showLCCN     \undefined \def \showLCCN      #1{\unskip}     \fi
\ifx \shownote     \undefined \def \shownote      #1{#1}          \fi
\ifx \showarticletitle \undefined \def \showarticletitle #1{#1}   \fi
\ifx \showURL      \undefined \def \showURL       {\relax}        \fi
% invisible to TeX
\providecommand\bibfield[2]{#2}
\providecommand\bibinfo[2]{#2}
\providecommand\natexlab[1]{#1}
\providecommand\showeprint[2][]{arXiv:#2}

\bibitem[\protect\citeauthoryear{Alizadeh, Edsall, Dharmapurikar, Vaidyanathan,
  Chu, Fingerhut, Lam, Matus, Pan, Yadav, et~al\mbox{.}}{Alizadeh
  et~al\mbox{.}}{2014}]%
        {alizadeh2014conga}
\bibfield{author}{\bibinfo{person}{Mohammad Alizadeh}, \bibinfo{person}{Tom
  Edsall}, \bibinfo{person}{Sarang Dharmapurikar}, \bibinfo{person}{Ramanan
  Vaidyanathan}, \bibinfo{person}{Kevin Chu}, \bibinfo{person}{Andy Fingerhut},
  \bibinfo{person}{Vinh~The Lam}, \bibinfo{person}{Francis Matus},
  \bibinfo{person}{Rong Pan}, \bibinfo{person}{Navindra Yadav},
  {et~al\mbox{.}}} \bibinfo{year}{2014}\natexlab{}.
\newblock \showarticletitle{CONGA: Distributed congestion-aware load balancing
  for datacenters}. In \bibinfo{booktitle}{\emph{Proceedings of the 2014 ACM
  conference on SIGCOMM}}. \bibinfo{pages}{503--514}.
\newblock


\bibitem[\protect\citeauthoryear{Amaro, Luo, Ousterhout, Krishnamurthy, Panda,
  Ratnasamy, and Shenker}{Amaro et~al\mbox{.}}{2020}]%
        {amaro2020remote}
\bibfield{author}{\bibinfo{person}{Emmanuel Amaro}, \bibinfo{person}{Zhihong
  Luo}, \bibinfo{person}{Amy Ousterhout}, \bibinfo{person}{Arvind
  Krishnamurthy}, \bibinfo{person}{Aurojit Panda}, \bibinfo{person}{Sylvia
  Ratnasamy}, {and} \bibinfo{person}{Scott Shenker}.}
  \bibinfo{year}{2020}\natexlab{}.
\newblock \showarticletitle{Remote Memory Calls}. In
  \bibinfo{booktitle}{\emph{Proceedings of the 19th ACM Workshop on Hot Topics
  in Networks}}. \bibinfo{pages}{38--44}.
\newblock


\bibitem[\protect\citeauthoryear{Arista}{Arista}{[n. d.]}]%
        {aristaTelemetry}
\bibfield{author}{\bibinfo{person}{Arista}.} \bibinfo{year}{[n.
  d.]}\natexlab{}.
\newblock \bibinfo{title}{Telemetry and Analytics}.
\newblock
  \bibinfo{howpublished}{\url{https://www.arista.com/en/solutions/telemetry-analytics}}.
    (\bibinfo{year}{[n. d.]}).
\newblock
\newblock
\shownote{Accessed: 2021-06-24.}


\bibitem[\protect\citeauthoryear{Association}{Association}{[n. d.]}]%
        {rocev2}
\bibfield{author}{\bibinfo{person}{Infiniband~Trade Association}.}
  \bibinfo{year}{[n. d.]}\natexlab{}.
\newblock \bibinfo{title}{RoCEv2}.
\newblock
  \bibinfo{howpublished}{\url{https://cw.infinibandta.org/document/dl/7781}}.
  (\bibinfo{year}{[n. d.]}).
\newblock
\newblock
\shownote{Accessed: 2021-05-12.}


\bibitem[\protect\citeauthoryear{Basat, Chen, Einziger, Feibish, Raz, and
  Yu}{Basat et~al\mbox{.}}{2020b}]%
        {basat2020routing}
\bibfield{author}{\bibinfo{person}{Ran~Ben Basat}, \bibinfo{person}{Xiaoqi
  Chen}, \bibinfo{person}{Gil Einziger}, \bibinfo{person}{Shir~Landau Feibish},
  \bibinfo{person}{Danny Raz}, {and} \bibinfo{person}{Minlan Yu}.}
  \bibinfo{year}{2020}\natexlab{b}.
\newblock \showarticletitle{Routing Oblivious Measurement Analytics}. In
  \bibinfo{booktitle}{\emph{2020 IFIP Networking Conference (Networking)}}.
  IEEE, \bibinfo{pages}{449--457}.
\newblock


\bibitem[\protect\citeauthoryear{Basat, Chen, Einziger, and
  Rottenstreich}{Basat et~al\mbox{.}}{2020a}]%
        {basat2020designing}
\bibfield{author}{\bibinfo{person}{Ran~Ben Basat}, \bibinfo{person}{Xiaoqi
  Chen}, \bibinfo{person}{Gil Einziger}, {and} \bibinfo{person}{Ori
  Rottenstreich}.} \bibinfo{year}{2020}\natexlab{a}.
\newblock \showarticletitle{Designing heavy-hitter detection algorithms for
  programmable switches}.
\newblock \bibinfo{journal}{\emph{IEEE/ACM Transactions on Networking}}
  \bibinfo{volume}{28}, \bibinfo{number}{3} (\bibinfo{year}{2020}),
  \bibinfo{pages}{1172--1185}.
\newblock


\bibitem[\protect\citeauthoryear{Ben~Basat, Ramanathan, Li, Antichi, Yu, and
  Mitzenmacher}{Ben~Basat et~al\mbox{.}}{2020}]%
        {ben2020pint}
\bibfield{author}{\bibinfo{person}{Ran Ben~Basat},
  \bibinfo{person}{Sivaramakrishnan Ramanathan}, \bibinfo{person}{Yuliang Li},
  \bibinfo{person}{Gianni Antichi}, \bibinfo{person}{Minian Yu}, {and}
  \bibinfo{person}{Michael Mitzenmacher}.} \bibinfo{year}{2020}\natexlab{}.
\newblock \showarticletitle{PINT: probabilistic in-band network telemetry}. In
  \bibinfo{booktitle}{\emph{Proceedings of the Annual conference of the ACM
  Special Interest Group on Data Communication on the applications,
  technologies, architectures, and protocols for computer communication}}.
  \bibinfo{pages}{662--680}.
\newblock


\bibitem[\protect\citeauthoryear{Bosshart, Daly, Gibb, Izzard, McKeown,
  Rexford, Schlesinger, Talayco, Vahdat, Varghese, et~al\mbox{.}}{Bosshart
  et~al\mbox{.}}{2014}]%
        {bosshart2014p4}
\bibfield{author}{\bibinfo{person}{Pat Bosshart}, \bibinfo{person}{Dan Daly},
  \bibinfo{person}{Glen Gibb}, \bibinfo{person}{Martin Izzard},
  \bibinfo{person}{Nick McKeown}, \bibinfo{person}{Jennifer Rexford},
  \bibinfo{person}{Cole Schlesinger}, \bibinfo{person}{Dan Talayco},
  \bibinfo{person}{Amin Vahdat}, \bibinfo{person}{George Varghese},
  {et~al\mbox{.}}} \bibinfo{year}{2014}\natexlab{}.
\newblock \showarticletitle{P4: Programming protocol-independent packet
  processors}.
\newblock \bibinfo{journal}{\emph{ACM SIGCOMM Computer Communication Review}}
  \bibinfo{volume}{44}, \bibinfo{number}{3} (\bibinfo{year}{2014}),
  \bibinfo{pages}{87--95}.
\newblock


\bibitem[\protect\citeauthoryear{BROADCOM}{BROADCOM}{[n. d.]}]%
        {trident}
\bibfield{author}{\bibinfo{person}{BROADCOM}.} \bibinfo{year}{[n.
  d.]}\natexlab{}.
\newblock \bibinfo{title}{{Trident Programmable Switch}}.
\newblock
  \bibinfo{howpublished}{\url{https://www.broadcom.com/products/ethernet-connectivity/switching/strataxgs/bcm56870-series}}.
    (\bibinfo{year}{[n. d.]}).
\newblock


\bibitem[\protect\citeauthoryear{Broder and Mitzenmacher}{Broder and
  Mitzenmacher}{2004}]%
        {broder2004network}
\bibfield{author}{\bibinfo{person}{Andrei Broder} {and}
  \bibinfo{person}{Michael Mitzenmacher}.} \bibinfo{year}{2004}\natexlab{}.
\newblock \showarticletitle{Network applications of bloom filters: A survey}.
\newblock \bibinfo{journal}{\emph{Internet mathematics}} \bibinfo{volume}{1},
  \bibinfo{number}{4} (\bibinfo{year}{2004}), \bibinfo{pages}{485--509}.
\newblock


\bibitem[\protect\citeauthoryear{Cisco}{Cisco}{[n. d.]}]%
        {ciscoMDT}
\bibfield{author}{\bibinfo{person}{Cisco}.} \bibinfo{year}{[n. d.]}\natexlab{}.
\newblock \bibinfo{title}{Explore Model-Driven Telemetry}.
\newblock
  \bibinfo{howpublished}{\url{https://blogs.cisco.com/developer/model-driven-telemetry-sandbox}}.
    (\bibinfo{year}{[n. d.]}).
\newblock
\newblock
\shownote{Accessed: 2021-06-24.}


\bibitem[\protect\citeauthoryear{Esmaeilzadeh, Blem, St.~Amant, Sankaralingam,
  and Burger}{Esmaeilzadeh et~al\mbox{.}}{2011}]%
        {esmaeilzadeh11}
\bibfield{author}{\bibinfo{person}{Hadi Esmaeilzadeh}, \bibinfo{person}{Emily
  Blem}, \bibinfo{person}{Renee St.~Amant}, \bibinfo{person}{Karthikeyan
  Sankaralingam}, {and} \bibinfo{person}{Doug Burger}.}
  \bibinfo{year}{2011}\natexlab{}.
\newblock \showarticletitle{{Dark silicon and the end of multicore scaling}}.
  In \bibinfo{booktitle}{\emph{International Symposium on Computer Architecture
  (ISCA)}}. \bibinfo{publisher}{ACM}.
\newblock


\bibitem[\protect\citeauthoryear{Garg}{Garg}{2013}]%
        {garg2013apache}
\bibfield{author}{\bibinfo{person}{Nishant Garg}.}
  \bibinfo{year}{2013}\natexlab{}.
\newblock \bibinfo{booktitle}{\emph{Apache kafka}}.
\newblock \bibinfo{publisher}{Packt Publishing Ltd}.
\newblock


\bibitem[\protect\citeauthoryear{Goodrich and Mitzenmacher}{Goodrich and
  Mitzenmacher}{2011}]%
        {goodrich2011invertible}
\bibfield{author}{\bibinfo{person}{Michael~T Goodrich} {and}
  \bibinfo{person}{Michael Mitzenmacher}.} \bibinfo{year}{2011}\natexlab{}.
\newblock \showarticletitle{Invertible bloom lookup tables}. In
  \bibinfo{booktitle}{\emph{2011 49th Annual Allerton Conference on
  Communication, Control, and Computing (Allerton)}}. IEEE,
  \bibinfo{pages}{792--799}.
\newblock


\bibitem[\protect\citeauthoryear{Group}{Group}{[n. d.]}]%
        {INTSpec}
\bibfield{author}{\bibinfo{person}{The P4.org Applications~Working Group}.}
  \bibinfo{year}{[n. d.]}\natexlab{}.
\newblock \bibinfo{title}{Telemetry Report Format Specification}.
\newblock
  \bibinfo{howpublished}{\url{https://github.com/p4lang/p4-applications/blob/master/docs/telemetry_report_latest.pdf}}.
    (\bibinfo{year}{[n. d.]}).
\newblock
\newblock
\shownote{Accessed: 2021-06-23.}


\bibitem[\protect\citeauthoryear{Guo, Yuan, Xiang, Dang, Huang, Maltz, Liu,
  Wang, Pang, Chen, et~al\mbox{.}}{Guo et~al\mbox{.}}{2015}]%
        {guo2015pingmesh}
\bibfield{author}{\bibinfo{person}{Chuanxiong Guo}, \bibinfo{person}{Lihua
  Yuan}, \bibinfo{person}{Dong Xiang}, \bibinfo{person}{Yingnong Dang},
  \bibinfo{person}{Ray Huang}, \bibinfo{person}{Dave Maltz},
  \bibinfo{person}{Zhaoyi Liu}, \bibinfo{person}{Vin Wang},
  \bibinfo{person}{Bin Pang}, \bibinfo{person}{Hua Chen}, {et~al\mbox{.}}}
  \bibinfo{year}{2015}\natexlab{}.
\newblock \showarticletitle{Pingmesh: A large-scale system for data center
  network latency measurement and analysis}. In
  \bibinfo{booktitle}{\emph{Proceedings of the 2015 ACM Conference on Special
  Interest Group on Data Communication}}. \bibinfo{pages}{139--152}.
\newblock


\bibitem[\protect\citeauthoryear{Gupta, Harrison, Canini, Feamster, Rexford,
  and Willinger}{Gupta et~al\mbox{.}}{2018}]%
        {gupta2018sonata}
\bibfield{author}{\bibinfo{person}{Arpit Gupta}, \bibinfo{person}{Rob
  Harrison}, \bibinfo{person}{Marco Canini}, \bibinfo{person}{Nick Feamster},
  \bibinfo{person}{Jennifer Rexford}, {and} \bibinfo{person}{Walter
  Willinger}.} \bibinfo{year}{2018}\natexlab{}.
\newblock \showarticletitle{Sonata: Query-driven streaming network telemetry}.
  In \bibinfo{booktitle}{\emph{Proceedings of the 2018 conference of the ACM
  special interest group on data communication}}. \bibinfo{pages}{357--371}.
\newblock


\bibitem[\protect\citeauthoryear{Handigol, Heller, Jeyakumar, Mazi\`{e}res, and
  McKeown}{Handigol et~al\mbox{.}}{2014}]%
        {handigol14}
\bibfield{author}{\bibinfo{person}{Nikhil Handigol}, \bibinfo{person}{Brandon
  Heller}, \bibinfo{person}{Vimalkumar Jeyakumar}, \bibinfo{person}{David
  Mazi\`{e}res}, {and} \bibinfo{person}{Nick McKeown}.}
  \bibinfo{year}{2014}\natexlab{}.
\newblock \showarticletitle{{I Know What Your Packet Did Last Hop: Using Packet
  Histories to Troubleshoot Networks}}. In \bibinfo{booktitle}{\emph{Networked
  Systems Design and Implementation (NSDI)}}. \bibinfo{publisher}{USENIX
  Association}.
\newblock


\bibitem[\protect\citeauthoryear{Hardavellas}{Hardavellas}{2012}]%
        {theRiseOfDarkSilicon}
\bibfield{author}{\bibinfo{person}{Nikos Hardavellas}.}
  \bibinfo{year}{2012}\natexlab{}.
\newblock \showarticletitle{{The rise and fall of dark silicon}}. In
  \bibinfo{booktitle}{\emph{;login:, Volume: 37}}. \bibinfo{publisher}{USENIX}.
\newblock


\bibitem[\protect\citeauthoryear{Hare}{Hare}{2011}]%
        {hare2011simple}
\bibfield{author}{\bibinfo{person}{Chris Hare}.}
  \bibinfo{year}{2011}\natexlab{}.
\newblock \bibinfo{title}{Simple Network Management Protocol (SNMP).}
\newblock   (\bibinfo{year}{2011}).
\newblock


\bibitem[\protect\citeauthoryear{Heller, Seetharaman, Mahadevan, Yiakoumis,
  Sharma, Banerjee, and McKeown}{Heller et~al\mbox{.}}{2010}]%
        {heller2010elastictree}
\bibfield{author}{\bibinfo{person}{Brandon Heller}, \bibinfo{person}{Srinivasan
  Seetharaman}, \bibinfo{person}{Priya Mahadevan}, \bibinfo{person}{Yiannis
  Yiakoumis}, \bibinfo{person}{Puneet Sharma}, \bibinfo{person}{Sujata
  Banerjee}, {and} \bibinfo{person}{Nick McKeown}.}
  \bibinfo{year}{2010}\natexlab{}.
\newblock \showarticletitle{Elastictree: Saving energy in data center
  networks.}. In \bibinfo{booktitle}{\emph{Nsdi}}, Vol.~\bibinfo{volume}{10}.
  \bibinfo{pages}{249--264}.
\newblock


\bibitem[\protect\citeauthoryear{Huang, Sun, Lee, Bai, Zhu, and Bao}{Huang
  et~al\mbox{.}}{2020}]%
        {huang2020omnimon}
\bibfield{author}{\bibinfo{person}{Qun Huang}, \bibinfo{person}{Haifeng Sun},
  \bibinfo{person}{Patrick~PC Lee}, \bibinfo{person}{Wei Bai},
  \bibinfo{person}{Feng Zhu}, {and} \bibinfo{person}{Yungang Bao}.}
  \bibinfo{year}{2020}\natexlab{}.
\newblock \showarticletitle{Omnimon: Re-architecting network telemetry with
  resource efficiency and full accuracy}. In
  \bibinfo{booktitle}{\emph{Proceedings of the Annual conference of the ACM
  Special Interest Group on Data Communication on the applications,
  technologies, architectures, and protocols for computer communication}}.
  \bibinfo{pages}{404--421}.
\newblock


\bibitem[\protect\citeauthoryear{Huawei}{Huawei}{[n. d.]}]%
        {huaweiTelemetry}
\bibfield{author}{\bibinfo{person}{Huawei}.} \bibinfo{year}{[n.
  d.]}\natexlab{}.
\newblock \bibinfo{title}{Overview of Telemetry}.
\newblock
  \bibinfo{howpublished}{\url{https://support.huawei.com/enterprise/en/doc/EDOC1000173015/165fa2c8/overview-of-telemetry}}.
    (\bibinfo{year}{[n. d.]}).
\newblock
\newblock
\shownote{Accessed: 2021-06-24.}


\bibitem[\protect\citeauthoryear{{Infiniband Trade Association}}{{Infiniband
  Trade Association}}{2015}]%
        {rdma}
\bibfield{author}{\bibinfo{person}{{Infiniband Trade Association}}.}
  \bibinfo{year}{2015}\natexlab{}.
\newblock \bibinfo{title}{InfiniBandTM Architecture Specification}.
\newblock   (\bibinfo{year}{2015}).
\newblock
\newblock
\shownote{Volume 1 Release 1.3.}


\bibitem[\protect\citeauthoryear{Intel}{Intel}{[n. d.]a}]%
        {intelINT}
\bibfield{author}{\bibinfo{person}{Intel}.} \bibinfo{year}{[n.
  d.]}\natexlab{a}.
\newblock \bibinfo{title}{In-band Network Telemetry Detects Network Performance
  Issues}.
\newblock
  \bibinfo{howpublished}{\url{https://builders.intel.com/docs/networkbuilders/in-band-network-telemetry-detects-network-performance-issues.pdf}}.
    (\bibinfo{year}{[n. d.]}).
\newblock
\newblock
\shownote{Accessed: 2021-06-04.}


\bibitem[\protect\citeauthoryear{Intel}{Intel}{[n. d.]b}]%
        {intelTofino}
\bibfield{author}{\bibinfo{person}{Intel}.} \bibinfo{year}{[n.
  d.]}\natexlab{b}.
\newblock \bibinfo{title}{Intel Tofino Series Programmable Ethernet Switch
  ASIC}.
\newblock
  \bibinfo{howpublished}{\url{https://www.intel.com/content/www/us/en/products/network-io/programmable-ethernet-switch/tofino-series/tofino.html}}.
    (\bibinfo{year}{[n. d.]}).
\newblock
\newblock
\shownote{Accessed: 2021-05-12.}


\bibitem[\protect\citeauthoryear{Intel}{Intel}{[n. d.]c}]%
        {intelDeepInsight}
\bibfield{author}{\bibinfo{person}{Intel}.} \bibinfo{year}{[n.
  d.]}\natexlab{c}.
\newblock \bibinfo{title}{Intel® Deep Insight Network Analytics Software}.
\newblock
  \bibinfo{howpublished}{\url{https://www.intel.com/content/www/us/en/products/network-io/programmable-ethernet-switch/network-analytics/deep-insight.html}}.
    (\bibinfo{year}{[n. d.]}).
\newblock
\newblock
\shownote{Accessed: 2021-06-10.}


\bibitem[\protect\citeauthoryear{Intel}{Intel}{[n. d.]d}]%
        {intelE810}
\bibfield{author}{\bibinfo{person}{Intel}.} \bibinfo{year}{[n.
  d.]}\natexlab{d}.
\newblock \bibinfo{title}{Intel® Ethernet Network Adapter E810-CQDA1/CQDA2}.
\newblock
  \bibinfo{howpublished}{\url{https://www.intel.com/content/www/us/en/products/docs/network-io/ethernet/network-adapters/ethernet-800-series-network-adapters/e810-cqda1-cqda2-100gbe-brief.html}}.
    (\bibinfo{year}{[n. d.]}).
\newblock
\newblock
\shownote{Accessed: 2021-06-11.}


\bibitem[\protect\citeauthoryear{Intel}{Intel}{[n. d.]e}]%
        {p4studio}
\bibfield{author}{\bibinfo{person}{Intel}.} \bibinfo{year}{[n.
  d.]}\natexlab{e}.
\newblock \bibinfo{title}{Intel® P4 Studio}.
\newblock
  \bibinfo{howpublished}{\url{https://www.intel.com/content/www/us/en/products/network-io/programmable-ethernet-switch/p4-suite/p4-studio.html}}.
    (\bibinfo{year}{[n. d.]}).
\newblock
\newblock
\shownote{Accessed: 2021-06-08.}


\bibitem[\protect\citeauthoryear{Jeyakumar, Alizadeh, Geng, Kim, and
  Mazi{\`e}res}{Jeyakumar et~al\mbox{.}}{2014}]%
        {jeyakumar2014millions}
\bibfield{author}{\bibinfo{person}{Vimalkumar Jeyakumar},
  \bibinfo{person}{Mohammad Alizadeh}, \bibinfo{person}{Yilong Geng},
  \bibinfo{person}{Changhoon Kim}, {and} \bibinfo{person}{David Mazi{\`e}res}.}
  \bibinfo{year}{2014}\natexlab{}.
\newblock \showarticletitle{Millions of little minions: Using packets for low
  latency network programming and visibility}.
\newblock \bibinfo{journal}{\emph{ACM SIGCOMM Computer Communication Review}}
  \bibinfo{volume}{44}, \bibinfo{number}{4} (\bibinfo{year}{2014}),
  \bibinfo{pages}{3--14}.
\newblock


\bibitem[\protect\citeauthoryear{Khandelwal, Agarwal, and Stoica}{Khandelwal
  et~al\mbox{.}}{2019}]%
        {khandelwal2019confluo}
\bibfield{author}{\bibinfo{person}{Anurag Khandelwal}, \bibinfo{person}{Rachit
  Agarwal}, {and} \bibinfo{person}{Ion Stoica}.}
  \bibinfo{year}{2019}\natexlab{}.
\newblock \showarticletitle{Confluo: Distributed monitoring and diagnosis stack
  for high-speed networks}. In \bibinfo{booktitle}{\emph{16th $\{$USENIX$\}$
  Symposium on Networked Systems Design and Implementation ($\{$NSDI$\}$ 19)}}.
  \bibinfo{pages}{421--436}.
\newblock


\bibitem[\protect\citeauthoryear{Kim, Sivaraman, Katta, Bas, Dixit, and
  Wobker}{Kim et~al\mbox{.}}{2015}]%
        {kim2015band}
\bibfield{author}{\bibinfo{person}{Changhoon Kim}, \bibinfo{person}{Anirudh
  Sivaraman}, \bibinfo{person}{Naga Katta}, \bibinfo{person}{Antonin Bas},
  \bibinfo{person}{Advait Dixit}, {and} \bibinfo{person}{Lawrence~J Wobker}.}
  \bibinfo{year}{2015}\natexlab{}.
\newblock \showarticletitle{In-band network telemetry via programmable
  dataplanes}. In \bibinfo{booktitle}{\emph{ACM SIGCOMM}}.
\newblock


\bibitem[\protect\citeauthoryear{Kim, Liu, Zhu, Kim, Lee, Sekar, and
  Seshan}{Kim et~al\mbox{.}}{2020}]%
        {kim2020tea}
\bibfield{author}{\bibinfo{person}{Daehyeok Kim}, \bibinfo{person}{Zaoxing
  Liu}, \bibinfo{person}{Yibo Zhu}, \bibinfo{person}{Changhoon Kim},
  \bibinfo{person}{Jeongkeun Lee}, \bibinfo{person}{Vyas Sekar}, {and}
  \bibinfo{person}{Srinivasan Seshan}.} \bibinfo{year}{2020}\natexlab{}.
\newblock \showarticletitle{Tea: Enabling state-intensive network functions on
  programmable switches}. In \bibinfo{booktitle}{\emph{Proceedings of the
  Annual conference of the ACM Special Interest Group on Data Communication on
  the applications, technologies, architectures, and protocols for computer
  communication}}. \bibinfo{pages}{90--106}.
\newblock


\bibitem[\protect\citeauthoryear{Ku{\v{c}}era, Popescu, Wang, Moore,
  Ko{\v{r}}enek, and Antichi}{Ku{\v{c}}era et~al\mbox{.}}{2020}]%
        {kuvcera2020enabling}
\bibfield{author}{\bibinfo{person}{Jan Ku{\v{c}}era},
  \bibinfo{person}{Diana~Andreea Popescu}, \bibinfo{person}{Han Wang},
  \bibinfo{person}{Andrew Moore}, \bibinfo{person}{Jan Ko{\v{r}}enek}, {and}
  \bibinfo{person}{Gianni Antichi}.} \bibinfo{year}{2020}\natexlab{}.
\newblock \showarticletitle{Enabling event-triggered data plane monitoring}. In
  \bibinfo{booktitle}{\emph{Proceedings of the Symposium on SDN Research}}.
  \bibinfo{pages}{14--26}.
\newblock


\bibitem[\protect\citeauthoryear{Li, Gao, Jin, and Xu}{Li
  et~al\mbox{.}}{2020}]%
        {li2020concerto}
\bibfield{author}{\bibinfo{person}{Yiran Li}, \bibinfo{person}{Kevin Gao},
  \bibinfo{person}{Xin Jin}, {and} \bibinfo{person}{Wei Xu}.}
  \bibinfo{year}{2020}\natexlab{}.
\newblock \showarticletitle{Concerto: cooperative network-wide telemetry with
  controllable error rate}. In \bibinfo{booktitle}{\emph{Proceedings of the
  11th ACM SIGOPS Asia-Pacific Workshop on Systems}}.
  \bibinfo{pages}{114--121}.
\newblock


\bibitem[\protect\citeauthoryear{Li, Miao, Liu, Zhuang, Feng, Tang, Cao, Zhang,
  Kelly, Alizadeh, et~al\mbox{.}}{Li et~al\mbox{.}}{2019}]%
        {li2019hpcc}
\bibfield{author}{\bibinfo{person}{Yuliang Li}, \bibinfo{person}{Rui Miao},
  \bibinfo{person}{Hongqiang~Harry Liu}, \bibinfo{person}{Yan Zhuang},
  \bibinfo{person}{Fei Feng}, \bibinfo{person}{Lingbo Tang},
  \bibinfo{person}{Zheng Cao}, \bibinfo{person}{Ming Zhang},
  \bibinfo{person}{Frank Kelly}, \bibinfo{person}{Mohammad Alizadeh},
  {et~al\mbox{.}}} \bibinfo{year}{2019}\natexlab{}.
\newblock \showarticletitle{HPCC: high precision congestion control}.
\newblock In \bibinfo{booktitle}{\emph{Proceedings of the ACM Special Interest
  Group on Data Communication}}. \bibinfo{pages}{44--58}.
\newblock


\bibitem[\protect\citeauthoryear{Lipton}{Lipton}{1994}]%
        {lipton1994new}
\bibfield{author}{\bibinfo{person}{Richard~J Lipton}.}
  \bibinfo{year}{1994}\natexlab{}.
\newblock \showarticletitle{A new approach to information theory}. In
  \bibinfo{booktitle}{\emph{Annual Symposium on Theoretical Aspects of Computer
  Science}}. Springer, \bibinfo{pages}{699--708}.
\newblock


\bibitem[\protect\citeauthoryear{Liu, Manousis, Vorsanger, Sekar, and
  Braverman}{Liu et~al\mbox{.}}{2016}]%
        {liu2016one}
\bibfield{author}{\bibinfo{person}{Zaoxing Liu}, \bibinfo{person}{Antonis
  Manousis}, \bibinfo{person}{Gregory Vorsanger}, \bibinfo{person}{Vyas Sekar},
  {and} \bibinfo{person}{Vladimir Braverman}.} \bibinfo{year}{2016}\natexlab{}.
\newblock \showarticletitle{One sketch to rule them all: Rethinking network
  flow monitoring with univmon}. In \bibinfo{booktitle}{\emph{Proceedings of
  the 2016 ACM SIGCOMM Conference}}. \bibinfo{pages}{101--114}.
\newblock


\bibitem[\protect\citeauthoryear{Mansour, Janvier, and Fajardo}{Mansour
  et~al\mbox{.}}{2019}]%
        {mansour2019fpga}
\bibfield{author}{\bibinfo{person}{Wassim Mansour}, \bibinfo{person}{Nicolas
  Janvier}, {and} \bibinfo{person}{Pablo Fajardo}.}
  \bibinfo{year}{2019}\natexlab{}.
\newblock \showarticletitle{FPGA implementation of RDMA-based data acquisition
  system over 100-Gb ethernet}.
\newblock \bibinfo{journal}{\emph{IEEE Transactions on Nuclear Science}}
  \bibinfo{volume}{66}, \bibinfo{number}{7} (\bibinfo{year}{2019}),
  \bibinfo{pages}{1138--1143}.
\newblock


\bibitem[\protect\citeauthoryear{Mitzenmacher and Upfal}{Mitzenmacher and
  Upfal}{2017}]%
        {mitzenmacher2017probability}
\bibfield{author}{\bibinfo{person}{Michael Mitzenmacher} {and}
  \bibinfo{person}{Eli Upfal}.} \bibinfo{year}{2017}\natexlab{}.
\newblock \bibinfo{booktitle}{\emph{Probability and computing: Randomization
  and probabilistic techniques in algorithms and data analysis}}.
\newblock \bibinfo{publisher}{Cambridge university press}.
\newblock


\bibitem[\protect\citeauthoryear{Networks}{Networks}{[n. d.]}]%
        {juniperTelemetry}
\bibfield{author}{\bibinfo{person}{Juniper Networks}.} \bibinfo{year}{[n.
  d.]}\natexlab{}.
\newblock \bibinfo{title}{Overview of the Junos Telemetry Interface}.
\newblock
  \bibinfo{howpublished}{\url{https://www.juniper.net/documentation/us/en/software/junos/interfaces-telemetry/topics/concept/junos-telemetry-interface-oveview.html}}.
    (\bibinfo{year}{[n. d.]}).
\newblock
\newblock
\shownote{Accessed: 2021-06-24.}


\bibitem[\protect\citeauthoryear{NVIDIA}{NVIDIA}{[n. d.]}]%
        {mellanox_spectrum}
\bibfield{author}{\bibinfo{person}{NVIDIA}.} \bibinfo{year}{[n.
  d.]}\natexlab{}.
\newblock \bibinfo{title}{{NVIDIA Mellanox Spectrum Switch}}.
\newblock
  \bibinfo{howpublished}{\url{https://www.mellanox.com/files/doc-2020/pb-spectrum-switch.pdf}}.
    (\bibinfo{year}{[n. d.]}).
\newblock


\bibitem[\protect\citeauthoryear{Rasley, Stephens, Dixon, Rozner, Felter,
  Agarwal, Carter, and Fonseca}{Rasley et~al\mbox{.}}{2014}]%
        {rasley2014planck}
\bibfield{author}{\bibinfo{person}{Jeff Rasley}, \bibinfo{person}{Brent
  Stephens}, \bibinfo{person}{Colin Dixon}, \bibinfo{person}{Eric Rozner},
  \bibinfo{person}{Wes Felter}, \bibinfo{person}{Kanak Agarwal},
  \bibinfo{person}{John Carter}, {and} \bibinfo{person}{Rodrigo Fonseca}.}
  \bibinfo{year}{2014}\natexlab{}.
\newblock \showarticletitle{Planck: Millisecond-scale monitoring and control
  for commodity networks}.
\newblock \bibinfo{journal}{\emph{ACM SIGCOMM Computer Communication Review}}
  \bibinfo{volume}{44}, \bibinfo{number}{4} (\bibinfo{year}{2014}),
  \bibinfo{pages}{407--418}.
\newblock


\bibitem[\protect\citeauthoryear{Roy, Zeng, Bagga, Porter, and Snoeren}{Roy
  et~al\mbox{.}}{2015}]%
        {roy2015inside}
\bibfield{author}{\bibinfo{person}{Arjun Roy}, \bibinfo{person}{Hongyi Zeng},
  \bibinfo{person}{Jasmeet Bagga}, \bibinfo{person}{George Porter}, {and}
  \bibinfo{person}{Alex~C Snoeren}.} \bibinfo{year}{2015}\natexlab{}.
\newblock \showarticletitle{Inside the social network's (datacenter) network}.
  In \bibinfo{booktitle}{\emph{Proceedings of the 2015 ACM Conference on
  Special Interest Group on Data Communication}}. \bibinfo{pages}{123--137}.
\newblock


\bibitem[\protect\citeauthoryear{Sidler, Wang, Chiosa, Kulkarni, and
  Alonso}{Sidler et~al\mbox{.}}{2020}]%
        {sidler2020strom}
\bibfield{author}{\bibinfo{person}{David Sidler}, \bibinfo{person}{Zeke Wang},
  \bibinfo{person}{Monica Chiosa}, \bibinfo{person}{Amit Kulkarni}, {and}
  \bibinfo{person}{Gustavo Alonso}.} \bibinfo{year}{2020}\natexlab{}.
\newblock \showarticletitle{StRoM: smart remote memory}. In
  \bibinfo{booktitle}{\emph{Proceedings of the Fifteenth European Conference on
  Computer Systems}}. \bibinfo{pages}{1--16}.
\newblock


\bibitem[\protect\citeauthoryear{Tammana, Agarwal, and Lee}{Tammana
  et~al\mbox{.}}{2018}]%
        {tammana2018distributed}
\bibfield{author}{\bibinfo{person}{Praveen Tammana}, \bibinfo{person}{Rachit
  Agarwal}, {and} \bibinfo{person}{Myungjin Lee}.}
  \bibinfo{year}{2018}\natexlab{}.
\newblock \showarticletitle{Distributed network monitoring and debugging with
  switchpointer}. In \bibinfo{booktitle}{\emph{15th $\{$USENIX$\}$ Symposium on
  Networked Systems Design and Implementation ($\{$NSDI$\}$ 18)}}.
  \bibinfo{pages}{453--456}.
\newblock


\bibitem[\protect\citeauthoryear{team}{team}{[n. d.]}]%
        {DPDKPerf}
\bibfield{author}{\bibinfo{person}{Intel DPDK~Validation team}.}
  \bibinfo{year}{[n. d.]}\natexlab{}.
\newblock \bibinfo{title}{DPDK Intel NIC Performance Report Release 20.11}.
\newblock
  \bibinfo{howpublished}{\url{https://fast.dpdk.org/doc/perf/DPDK_20_11_Intel_NIC_performance_report.pdf}}.
    (\bibinfo{year}{[n. d.]}).
\newblock
\newblock
\shownote{Accessed: 2021-05-07.}


\bibitem[\protect\citeauthoryear{Technologies}{Technologies}{[n. d.]}]%
        {mellanoxConnectx6}
\bibfield{author}{\bibinfo{person}{Mellanox Technologies}.} \bibinfo{year}{[n.
  d.]}\natexlab{}.
\newblock \bibinfo{title}{ConnectX®-6 VPI Card}.
\newblock
  \bibinfo{howpublished}{\url{https://www.mellanox.com/files/doc-2020/pb-connectx-6-vpi-card.pdf}}.
    (\bibinfo{year}{[n. d.]}).
\newblock
\newblock
\shownote{Accessed: 2021-05-12.}


\bibitem[\protect\citeauthoryear{Van~Tu, Hyun, and Hong}{Van~Tu
  et~al\mbox{.}}{2017}]%
        {van2017towards}
\bibfield{author}{\bibinfo{person}{Nguyen Van~Tu}, \bibinfo{person}{Jonghwan
  Hyun}, {and} \bibinfo{person}{James Won-Ki Hong}.}
  \bibinfo{year}{2017}\natexlab{}.
\newblock \showarticletitle{Towards onos-based sdn monitoring using in-band
  network telemetry}. In \bibinfo{booktitle}{\emph{2017 19th Asia-Pacific
  Network Operations and Management Symposium (APNOMS)}}. IEEE,
  \bibinfo{pages}{76--81}.
\newblock


\bibitem[\protect\citeauthoryear{Van~Tu, Hyun, Kim, Yoo, and Hong}{Van~Tu
  et~al\mbox{.}}{2018}]%
        {van2018intcollector}
\bibfield{author}{\bibinfo{person}{Nguyen Van~Tu}, \bibinfo{person}{Jonghwan
  Hyun}, \bibinfo{person}{Ga~Yeon Kim}, \bibinfo{person}{Jae-Hyoung Yoo}, {and}
  \bibinfo{person}{James Won-Ki Hong}.} \bibinfo{year}{2018}\natexlab{}.
\newblock \showarticletitle{Intcollector: A high-performance collector for
  in-band network telemetry}. In \bibinfo{booktitle}{\emph{2018 14th
  International Conference on Network and Service Management (CNSM)}}. IEEE,
  \bibinfo{pages}{10--18}.
\newblock


\bibitem[\protect\citeauthoryear{Vestin, Kassler, Bhamare, Grinnemo, Andersson,
  and Pongracz}{Vestin et~al\mbox{.}}{2019}]%
        {vestin2019programmable}
\bibfield{author}{\bibinfo{person}{Jonathan Vestin}, \bibinfo{person}{Andreas
  Kassler}, \bibinfo{person}{Deval Bhamare}, \bibinfo{person}{Karl-Johan
  Grinnemo}, \bibinfo{person}{Jan-Olof Andersson}, {and}
  \bibinfo{person}{Gergely Pongracz}.} \bibinfo{year}{2019}\natexlab{}.
\newblock \showarticletitle{Programmable event detection for in-band network
  telemetry}. In \bibinfo{booktitle}{\emph{2019 IEEE 8th international
  conference on cloud networking (CloudNet)}}. IEEE, \bibinfo{pages}{1--6}.
\newblock


\bibitem[\protect\citeauthoryear{Xilinx}{Xilinx}{[n. d.]}]%
        {xilinxERNIC}
\bibfield{author}{\bibinfo{person}{Xilinx}.} \bibinfo{year}{[n.
  d.]}\natexlab{}.
\newblock \bibinfo{title}{Xilinx Embedded RDMA Enabled NIC}.
\newblock
  \bibinfo{howpublished}{\url{https://www.xilinx.com/support/documentation/ip_documentation/ernic/v3_0/pg332-ernic.pdf}}.
    (\bibinfo{year}{[n. d.]}).
\newblock
\newblock
\shownote{Accessed: 2021-06-11.}


\bibitem[\protect\citeauthoryear{Yu, Zhu, Arzani, Fonseca, Zhang, Deng, and
  Yuan}{Yu et~al\mbox{.}}{2019}]%
        {yu2019dshark}
\bibfield{author}{\bibinfo{person}{Da Yu}, \bibinfo{person}{Yibo Zhu},
  \bibinfo{person}{Behnaz Arzani}, \bibinfo{person}{Rodrigo Fonseca},
  \bibinfo{person}{Tianrong Zhang}, \bibinfo{person}{Karl Deng}, {and}
  \bibinfo{person}{Lihua Yuan}.} \bibinfo{year}{2019}\natexlab{}.
\newblock \showarticletitle{dShark: A general, easy to program and scalable
  framework for analyzing in-network packet traces}. In
  \bibinfo{booktitle}{\emph{16th $\{$USENIX$\}$ Symposium on Networked Systems
  Design and Implementation ($\{$NSDI$\}$ 19)}}. \bibinfo{pages}{207--220}.
\newblock


\bibitem[\protect\citeauthoryear{Yu}{Yu}{2019}]%
        {yu2019network}
\bibfield{author}{\bibinfo{person}{Minlan Yu}.}
  \bibinfo{year}{2019}\natexlab{}.
\newblock \showarticletitle{Network telemetry: towards a top-down approach}.
\newblock \bibinfo{journal}{\emph{ACM SIGCOMM Computer Communication Review}}
  \bibinfo{volume}{49}, \bibinfo{number}{1} (\bibinfo{year}{2019}),
  \bibinfo{pages}{11--17}.
\newblock


\bibitem[\protect\citeauthoryear{Zhou, Bi, Yang, Gao, Cao, Zhang, Wang, and
  Zhang}{Zhou et~al\mbox{.}}{2020a}]%
        {zhou2020hypersight}
\bibfield{author}{\bibinfo{person}{Yu Zhou}, \bibinfo{person}{Jun Bi},
  \bibinfo{person}{Tong Yang}, \bibinfo{person}{Kai Gao},
  \bibinfo{person}{Jiamin Cao}, \bibinfo{person}{Dai Zhang},
  \bibinfo{person}{Yangyang Wang}, {and} \bibinfo{person}{Cheng Zhang}.}
  \bibinfo{year}{2020}\natexlab{a}.
\newblock \showarticletitle{Hypersight: Towards scalable, high-coverage, and
  dynamic network monitoring queries}.
\newblock \bibinfo{journal}{\emph{IEEE Journal on Selected Areas in
  Communications}} \bibinfo{volume}{38}, \bibinfo{number}{6}
  (\bibinfo{year}{2020}), \bibinfo{pages}{1147--1160}.
\newblock


\bibitem[\protect\citeauthoryear{Zhou, Sun, Liu, Miao, Bai, Li, Zheng, Zhu,
  Shen, Xi, et~al\mbox{.}}{Zhou et~al\mbox{.}}{2020b}]%
        {zhou2020flow}
\bibfield{author}{\bibinfo{person}{Yu Zhou}, \bibinfo{person}{Chen Sun},
  \bibinfo{person}{Hongqiang~Harry Liu}, \bibinfo{person}{Rui Miao},
  \bibinfo{person}{Shi Bai}, \bibinfo{person}{Bo Li}, \bibinfo{person}{Zhilong
  Zheng}, \bibinfo{person}{Lingjun Zhu}, \bibinfo{person}{Zhen Shen},
  \bibinfo{person}{Yongqing Xi}, {et~al\mbox{.}}}
  \bibinfo{year}{2020}\natexlab{b}.
\newblock \showarticletitle{Flow event telemetry on programmable data plane}.
  In \bibinfo{booktitle}{\emph{Proceedings of the Annual conference of the ACM
  Special Interest Group on Data Communication on the applications,
  technologies, architectures, and protocols for computer communication}}.
  \bibinfo{pages}{76--89}.
\newblock


\bibitem[\protect\citeauthoryear{Zhu, Kang, Cao, Greenberg, Lu, Mahajan, Maltz,
  Yuan, Zhang, Zhao, et~al\mbox{.}}{Zhu et~al\mbox{.}}{2015}]%
        {zhu2015packet}
\bibfield{author}{\bibinfo{person}{Yibo Zhu}, \bibinfo{person}{Nanxi Kang},
  \bibinfo{person}{Jiaxin Cao}, \bibinfo{person}{Albert Greenberg},
  \bibinfo{person}{Guohan Lu}, \bibinfo{person}{Ratul Mahajan},
  \bibinfo{person}{Dave Maltz}, \bibinfo{person}{Lihua Yuan},
  \bibinfo{person}{Ming Zhang}, \bibinfo{person}{Ben~Y Zhao}, {et~al\mbox{.}}}
  \bibinfo{year}{2015}\natexlab{}.
\newblock \showarticletitle{Packet-level telemetry in large datacenter
  networks}. In \bibinfo{booktitle}{\emph{Proceedings of the 2015 ACM
  Conference on Special Interest Group on Data Communication}}.
  \bibinfo{pages}{479--491}.
\newblock


\end{thebibliography}

\end{document}